# Benchmarks and results of the two-band Hubbard model from the Gutzwiller conjugate gradient minimization theory

Zhuo Ye,[1,*] Feng Zhang,[1,2,*] Yong-Xin Yao,[1,2] Cai-Zhuang Wang,[1,2] Kai-Ming Ho[2]

[1] Ames National Laboratory – US DOE, Ames, Iowa 50011, United States

[2] Department of Physics and Astronomy, Iowa State University, Ames, Iowa 50011, United States

* E-mail: zye@iastate.edu (Z. Ye), fzhang@ameslab.gov (F. Zhang)

ABSTRACT

Ground-state properties, such as energies and double occupancies, of a one-dimensional two-band Hubbard model are calculated using a first principles Gutzwiller conjugate gradient minimization theory. The favorable agreement with the results from the density matrix renormalization group theory demonstrates the accuracy of our method. A rotationally invariant approach is further incorporated into the method to greatly reduce the computational complexity with a speedup of $50\times$. Moreover, we investigate the Mott transition between a metal and a Mott insulator by evaluating the charge gap. With greatly reduced computational effort, our method reproduces the phase diagram in reasonable agreement with the density matrix renormalization group theory.

1. INTRODUCTION

Accurate description of a large number of interacting electrons remains one of the grand scientific challenges of the present day even with the computer industry's explosive technology growth. The main source of difficulties lies in the fact that the dimension of the Hilbert space required to describe a correlated electron system grows exponentially in the system size and that the minus sign problem arising from strongly interacting fermions leads to very slow convergence of straightforward Monte Carlo approaches. Numerical studies of most large, realistic systems are affordable only via the Kohn–Sham density function theory (DFT) [1,2]. However, DFT often fails in describing strongly-correlated electron materials, where the behavior of electrons cannot be described effectively in terms of non-interacting entities. On the other hand, quantum chemistry approaches that are capable of accurately describing smaller systems are often too expensive for many large, realistic systems. Hybrid approaches that merge DFT with many-body techniques [3-5] have been demonstrated to be powerful and efficient in describing the properties of real correlated-electron materials. However, the use of adjustable screened Coulomb parameters restricts the predictive power of these methods. In order to study realistic condensed matter systems and predict unusual properties that emerge from the correlated behavior of electrons, it is highly desirable to develop *ab initio* methods that are both affordable and reasonably accurate.

We have been developing such a many-body approach, namely, the Gutzwiller conjugate gradient minimization (GCGM) method [6-12]. The GCGM method is based on the Gutzwiller wave function (GWF) that was proposed by Gutzwiller in the 1960s [13-15]. The GWF is a simple variational wave function which can be defined as a correlation projector acting on a one-particle product state. In GWF-based theories, the Gutzwiller approximation (GA, also proposed by Gutzwiller) is commonly adopted to evaluate the expectation value of an observable with respect to the GWF. However, this approximation 'decouples' the two correlated sites by using a site–site factorization and overlooks the correlation between electrons of parallel spins (the exchange hole) [16]. It recovers the exact GWF results only in infinite dimensions [17]. In finite dimensions it could be a major source of the inaccuracy although the use of the approximation greatly improves the computational efficiency [18]. In the GCGM theory, we use the GWF without using the GA, so that the expectation values observables can be evaluated in a more rigorous way. Secondly, the conventional GWF introduces correlations into the trial wave function $|\Psi_0\rangle$ only via an on-site correlation factor; while GCGM method reinforces a site–site constraint to account for inter-site correlations, which has been demonstrated to successful describe single-band Hubbard models in the large $U$ limit [10]. Our previous work has demonstrated the improved accuracy of the GCGM method with benchmark tests of molecules, such as energy calculations of both the ground and the excited states of dimers [7-9]. The benchmark tests on periodic bulk systems have focused mainly on the Hubbard model, both one- and two-dimensional, but with only one band [10]. Since our ultimate goal is to develop a numerical tool to access realistic condensed matter systems, the next benchmark test is naturally the multi-band Hubbard model. In this work, we focus on a general two-band Hubbard model and compare our results with those given by density matrix renormalization group theory (DMRG) [19,20], a well-established and powerful many-body approach, especially for one-dimensional (1D) systems. We examine the ground state properties, including the energy and the double occupancy, as well as the metal-Mott insulator-transition in these systems.

When dealing with multi-band/multi-orbital systems, a grand challenge is that the computational complexity grows exponentially with the number of orbitals. In the 2$^{nd}$ part, we introduce a rotationally invariant (RI) scheme to greatly reduce the computational cost. In the RI scheme, orbitals that satisfy certain symmetries can be grouped together and the Gutzwiller projector is parametrized based on the number of electrons occupying the group of orbitals instead of the onsite configurations. For example, for a cubic system of which the valence orbitals contain $p-$orbitals, we may group the $p_x, p_y, p_z$ orbitals (if they overlap with each other after rotation by 90° about one of the $x, y, z$ axes) and use the number of electrons occupying any of the $p-$orbitals to parametrize the system. As demonstrated with benchmark tests of dimers [7,12], the scheme efficiently groups the onsite orbitals according to their symmetry and greatly reduces the computational complexity without sacrificing accuracy. In the 2$^{nd}$ part of this work, we apply the RI scheme to the same 2-band Hubbard model and repeat the calculations of the 1$^{st}$ part. A speedup of $\sim 300\times$ associated with a favorable agreement with the results given by the original GCGM without applying RI demonstrates the efficiency and the accuracy of the RI scheme. We summarize this work by investigating a Mott transition in this system with GCGM+RI and comparing our results with DMRG.

## 2. MODELS AND METHODS

### 2.1. The 2-band Hubbard model

The Hubbard model is defined by the Hamiltonian that contains a kinetic energy term and a Coulomb interaction term,

$$\hat{H} = \hat{H}_K + \hat{H}_{Coul}, \tag{1}$$

where the kinetic term $\hat{H}_K$ is defined as,

$$\hat{H}_K = -t \sum_{\langle IJ \rangle \alpha \sigma} c^\dagger_{I\alpha\sigma} c_{J\alpha\sigma} - \left( t' \sum_{I\sigma} c^\dagger_{I1\sigma} c_{I2\sigma} + h.c. \right), \tag{2}$$

and the Coulomb interaction term $\hat{H}_{Coul}$ is defined as,

$$\hat{H}_{Coul} = U\sum_{I\alpha} n_{I\alpha\uparrow}n_{I\alpha\downarrow} + U'\sum_{I\sigma} n_{I1\sigma}n_{I2\bar{\sigma}} + (U'-J)\sum_{I\sigma} n_{I1\sigma}n_{I2\sigma}$$
$$+ J\sum_{I}\left(c^{\dagger}_{I1\uparrow}c^{\dagger}_{I2\downarrow}c_{I1\downarrow}c_{I2\uparrow} + c^{\dagger}_{I1\uparrow}c^{\dagger}_{I1\downarrow}c_{I2\downarrow}c_{I2\uparrow} + H.c.\right). \quad (3)$$

The operator $c^{\dagger}_{I\alpha\sigma}$ ($c_{I\alpha\sigma}$) creates (annihilates) an electron with spin $\sigma(=\uparrow,\downarrow)$ and orbital index $\alpha$ ($=1,2$) at the $I$ th site (or unit cell) and $n_{I\alpha\sigma} = c^{\dagger}_{I\alpha\sigma}c_{I\alpha\sigma}$. $t$ represents the orbital-independent nearest-neighbor hopping and $t'$ the hybridization between the two orbitals. $U$ ($U'$) represents the intra-band (inter-band) Coulomb interaction and $J$ the Hund coupling. $\bar{\sigma}$ denotes the opposite spin of $\sigma$, and $\langle IJ \rangle$ a pair of nearest neighbors. The standard constraint $U'=U-2J$ is used given by the cubic symmetry. We only consider the regime $0 \leq J \leq U/3$ throughout this study to ensure that all the Coulomb interaction coefficients in Eq. (3) are non-negative.

Under the linear transformation $\tilde{c}_{I1\sigma} = \frac{1}{\sqrt{2}}(c_{I1\sigma} + c_{I2\sigma})$ and $\tilde{c}_{I2\sigma} = \frac{1}{\sqrt{2}}(c_{I1\sigma} - c_{I2\sigma})$, where $\tilde{c}_{I1\sigma}/\tilde{c}_{I2\sigma}$ annihilates an electron from the bonding/anti-bonding band, respectively, the kinetic Hamiltonian term can be written in a more familiar form,

$$\hat{H}_{K} = \sum_{I\alpha\sigma} \Delta_{\alpha}\tilde{c}^{\dagger}_{I\alpha\sigma}\tilde{c}_{I\alpha\sigma} - t\sum_{\langle IJ \rangle\alpha\sigma} \tilde{c}^{\dagger}_{I\alpha\sigma}\tilde{c}_{J\alpha\sigma}, \quad (4)$$

where $\Delta_{\alpha}$ is the orbital-dependent crystal-field splitting and $\Delta_1 = -t'$, $\Delta_2 = t'$, and the Coulomb interaction term

$$\hat{H}_{Coul} = U\sum_{I\alpha} \tilde{n}_{I\alpha\uparrow}\tilde{n}_{I\alpha\downarrow} + U'\sum_{I\sigma} \tilde{n}_{I1\sigma}\tilde{n}_{I2\bar{\sigma}} + (U'-J)\sum_{I\sigma} \tilde{n}_{I1\sigma}\tilde{n}_{I2\sigma}$$
$$+ J\sum_{I}\left(\tilde{c}^{\dagger}_{I1\uparrow}\tilde{c}^{\dagger}_{I2\downarrow}\tilde{c}_{I1\downarrow}\tilde{c}_{I2\uparrow} + \tilde{c}^{\dagger}_{I1\uparrow}\tilde{c}^{\dagger}_{I1\downarrow}\tilde{c}_{I2\downarrow}\tilde{c}_{I2\uparrow} + H.c.\right), \quad (5)$$

where $\tilde{n}_{I\alpha\sigma} = \tilde{c}^{\dagger}_{I\alpha\sigma}\tilde{c}_{I\alpha\sigma}$. It can be seen that the Kanamori coulomb interaction term $\hat{H}_{Coul}$ keeps the same form as in Eq. (3) under the linear transformation, i.e., $\hat{H}_{Coul}$ is rotationally invariant.

2.2. The GCGM method

The formalism of the GCGM method for periodic systems has been presented elsewhere [6,7,10]. However, to facilitate a better understanding of the method, we still outline the essential formalism here. The formalism presented here has a general form and can be applied to any periodic system other than the Hubbard model. We start from a general nonrelativistic Hamiltonian for a periodic system,

$$\hat{H} = \sum_{Ii\alpha,Jj\beta,\sigma} t_{Ii\alpha Jj\beta} c^{\dagger}_{Ii\alpha\sigma} c_{Jj\beta\sigma} + \frac{1}{2} \sum_{\substack{Ii\alpha,Jj\beta \\ Kk\gamma,Ll\delta,\sigma\sigma'}} u(Ii\alpha, Jj\beta; Kk\gamma, Ll\delta) c^{\dagger}_{Ii\alpha\sigma} c^{\dagger}_{Jj\beta\sigma'} c_{Ll\delta\sigma'} c_{Kk\gamma\sigma}, \qquad (6)$$

where $I,J,K,L$ are the unit cell indices, $i,j,k,l$ the atomic site indices, $\alpha,\beta,\gamma,\delta$ the orbital indices, and $\sigma,\sigma'$ the spin indices. Here, $t$ and $u$ are the one-electron hopping integral and the two-electron Coulomb integral, respectively. We note that in the Hubbard model described in Eq. (1)-(5), there is only 1 atom per unit cell, and that the atomic site indices are therefore removed from these equations. The GWF is constructed from a trial wave function $|\Psi_0\rangle$ that is often non-interacting, i.e., a single Slater determinant, and can be written as,

$$|\Psi_{GWF}\rangle = \prod_{Ii}\left(\sum_{\Gamma} g(\Gamma_{Ii})|\Gamma_{Ii}\rangle\langle\Gamma_{Ii}|\right)|\Psi_0\rangle. \qquad (7)$$

Since all unit cells are identical, $g(\Gamma_{Ii})$ is independent on a specific unit cell and so $g(\Gamma_{Ii}) = g(\Gamma_i)$. The total energy of the system can be expressed as,

$$E_{GWF} = \sum_{Ii\alpha, Jj\beta, \sigma} t_{Ii\alpha, Jj\beta} \langle c^\dagger_{Ii\alpha\sigma} c_{Jj\beta\sigma} \rangle_{GWF} + \frac{1}{2} \sum_{\substack{Ii\alpha, Jj\beta \\ Kk\gamma, Ll\delta, \sigma\sigma'}} u(Ii\alpha, Jj\beta; Kk\gamma, Ll\delta) \langle c^\dagger_{Ii\alpha\sigma} c^\dagger_{Jj\beta\sigma'} c_{Ll\delta\sigma'} c_{Kk\gamma\sigma} \rangle_{GWF},$$

(8)

where $\langle \hat{O} \rangle_{GWF}$ is a short-hand notation for $\langle \Psi_{GWF} | \hat{O} | \Psi_{GWF} \rangle$. The 1-particle density matrix (1PDM) can be expressed as,

$$\langle c^\dagger_{Ii\alpha\sigma} c_{Ii\beta\sigma} \rangle_{GWF} = \frac{1}{\langle \Psi_{GWF} | \Psi_{GWF} \rangle_{Ii,Jj}} \sum_{\Gamma_{Ii}, \Gamma'_{Ii}, \Gamma_{Jj}} \langle \Gamma_{Ii} | c^\dagger_{Ii\alpha\sigma} c_{Ii\beta\sigma} | \Gamma'_{Ii} \rangle \cdot$$
$$g(\Gamma_i) g(\Gamma'_i) g(\Gamma_j)^2 \xi_{\Gamma_{Ii}, \Gamma_{Jj}, \Gamma'_{Ii}, \Gamma_{Jj}},$$

(9)

$$\langle c^\dagger_{Ii\alpha\sigma} c_{Jj\beta\sigma} \rangle_{GWF} = \frac{1}{\langle \Psi_{GWF} | \Psi_{GWF} \rangle_{Ii,Jj}} \sum_{\Gamma_{Ii}, \Gamma_{Jj}, \Gamma'_{Ii}, \Gamma'_{Jj}} \langle \Gamma_{Ii} | c^\dagger_{Ii\alpha\sigma} | \Gamma'_{Ii} \rangle \langle \Gamma_{Jj} | c_{Jj\beta\sigma} | \Gamma'_{Jj} \rangle \cdot$$
$$g(\Gamma_i) g(\Gamma_j) g(\Gamma'_i) g(\Gamma'_j) \xi_{\Gamma_{Ii}, \Gamma_{Jj}, \Gamma'_{Ii}, \Gamma'_{Jj}}, \text{ for } (I,i) \neq (J,j)$$

(10)

where the atomic site $Jj$ is chosen to be the nearest neighbor of site $Ii$ in Eq. (9), and $\xi_{\Gamma_{Ii}, \Gamma_{Jj}, \Gamma'_{Ii}, \Gamma'_{Jj}}$ is the coefficient determined from $|\Psi_0\rangle$ and Gutzwiller variational parameters,

$$\xi_{\Gamma_{Ii}, \Gamma_{Jj}, \Gamma'_{Ii}, \Gamma'_{Jj}} = \sum_{\{\Gamma_{Kk}, Kk \neq Ii, Jj\}} \prod_{Kk} g(\Gamma_k)^2 \langle \Psi_0 | \Gamma_{Ii}, \Gamma_{Jj}, \{\Gamma_{Kk}\} \rangle \langle \Gamma'_{Ii}, \Gamma'_{Jj}, \{\Gamma_{Kk}\} | \Psi_0 \rangle,$$

(11)

and

$$\langle \Psi_{GWF} | \Psi_{GWF} \rangle_{Ii,Jj} = \sum_{\Gamma_{Ii}, \Gamma_{Jj}} \xi_{\Gamma_{Ii}, \Gamma_{Jj}, \Gamma_{Ii}, \Gamma_{Jj}} g(\Gamma_i)^2 g(\Gamma_j)^2,$$

(12)

The on-site 2-particle correlation matrix (2PCM) can be expressed as,

$$\left\langle c^\dagger_{Ii\alpha\sigma} c^\dagger_{Ii\beta\sigma} c_{Ii\delta\sigma} c_{Ii\gamma\sigma} \right\rangle_{GWF} = \frac{1}{\left\langle \Psi_{GWF} | \Psi_{GWF} \right\rangle_{Ii,Jj}} \sum_{\Gamma_{Ii},\Gamma'_{Ii},\Gamma_{Jj}} \left\langle \Gamma_{Ii} \left| c^\dagger_{Ii\alpha\sigma} c^\dagger_{Ii\beta\sigma} c_{Ii\delta\sigma} c_{Ii\gamma\sigma} \right| \Gamma'_{Ii} \right\rangle \cdot$$
$$g(\Gamma_i) g(\Gamma'_i) g(\Gamma_j)^2 \xi_{\Gamma_{Ii},\Gamma_{Jj},\Gamma'_{Ii},\Gamma_{Jj}}, \quad (13)$$

where the atomic site $Jj$ is chosen to be the nearest neighbor of site $Ii$. For the Hubbard model discussed in this work, only the onsite 2-particle correlation matrix (2PCM) $\left\langle c^\dagger_{Ii\alpha\sigma} c^\dagger_{Ii\beta\sigma} c_{Ii\delta\sigma} c_{Ii\gamma\sigma} \right\rangle_{GWF}$ are included, which can be evaluated using Eq. (13). For a more general system, we can still use Eq. (13) to evaluate the onsite 2PCM and use Wick's theorem to evaluate approximately the inter-site 2PCM $\left\langle c^\dagger_{Ii\alpha\sigma} c^\dagger_{Jj\beta\sigma} c_{Ll\delta\sigma} c_{Kk\gamma\sigma} \right\rangle_{GWF}$ when $(Ii, Jj, Kk, Ll)$ are not all equal [7].

The expression in Eq. (9)-(13) are accurate within the GWF framework. However, as $\xi_{\Gamma_{Ii},\Gamma_{Jj},\Gamma'_{Ii},\Gamma'_{Jj}}$ is determined from both $\left| \Psi_0 \right\rangle$ and the Gutzwiller variational parameters of all the rest sites other than $(Ii, Jj)$, it is a grand challenge to rigorously evaluate the coefficient $\xi_{\Gamma_{Ii},\Gamma_{Jj},\Gamma'_{Ii},\Gamma'_{Jj}}$ because the computational complexity grows exponentially with respect to the number of atomic sites. Therefore, we use an effective approximation to evaluate $\xi_{\Gamma_{Ii},\Gamma_{Jj},\Gamma'_{Ii},\Gamma'_{Jj}}$ from $\xi^0_{\Gamma_{Ii},\Gamma_{Jj},\Gamma'_{Ii},\Gamma'_{Jj}}$ defined as follows,

$$\begin{aligned}
\xi^0_{\Gamma_{Ii},\Gamma_{Jj},\Gamma'_{Ii},\Gamma'_{Jj}} &= \left\langle \Psi_0 | \Gamma_{Ii},\Gamma_{Jj} \right\rangle \left\langle \Gamma'_{Ii},\Gamma'_{Jj} | \Psi_0 \right\rangle \\
&= \sum_{\{\Gamma_{Kk}, Kk \neq Ii, Jj\}} \left\langle \Psi_0 | \Gamma_{Ii},\Gamma_{Jj},\{\Gamma_{Kk}\} \right\rangle \left\langle \Gamma'_{Ii},\Gamma'_{Jj},\{\Gamma_{Kk}\} | \Psi_0 \right\rangle \\
&= \left\langle \Psi_0 \left| c^\dagger_{Ii}(\Gamma_{Ii}) c_{Ii}(\Gamma'_{Ii}) c^\dagger_{Jj}(\Gamma_{Jj}) c_{Jj}(\Gamma'_{Jj}) \right| \Psi_0 \right\rangle.
\end{aligned} \quad (14)$$

Then $\xi_{\Gamma_{Ii},\Gamma_{Jj},\Gamma'_{Ii},\Gamma'_{Jj}}$ can be estimated approximately using the following expression,

$$\xi_{\Gamma_{Ii},\Gamma_{Jj},\Gamma'_{Ii},\Gamma'_{Jj}} \approx \xi^0_{\Gamma_{Ii},\Gamma_{Jj},\Gamma'_{Ii},\Gamma'_{Jj}} F_\uparrow\left(n_{Ii\uparrow}, n_{Jj\uparrow}\right) F_\downarrow\left(n_{Ii\downarrow}, n_{Jj\downarrow}\right) \quad (15)$$

where

$$F_\sigma\left(n_{Ii\sigma},n_{Jj\sigma}\right)=\begin{cases}\left(\sum_{\Gamma_k}p(\Gamma_k)\dfrac{g(\Gamma_k^{\sigma+})^2}{g(\Gamma_k)^2}\right)^{-(n_{Ii\sigma}+n_{Jj\sigma}-n_{0\sigma})} & n_{Ii\sigma}+n_{Jj\sigma}<n_{0\sigma} \\ 1 & n_{Ii\sigma}+n_{Jj\sigma}=n_{0\sigma} \\ \left(\sum_{\Gamma_k}p(\Gamma_k)\dfrac{g(\Gamma_k^{\sigma-})^2}{g(\Gamma_k)^2}\right)^{n_{Ii\sigma}+n_{Jj\sigma}-n_{0\sigma}} & n_{Ii\sigma}+n_{Jj\sigma}>n_{0\sigma},\end{cases} \quad (16)$$

where $n_{Ii\sigma}$ and $n_{Jj\sigma}$ are the number of electrons with spin $\sigma$ occupying site $Ii$ and $Jj$, respectively, and $n_0$ is the expected total number of electrons with spin $\sigma$ occupying site $Ii$ and $Jj$. $p(\Gamma_k)$ is the probability of finding the configuration $\Gamma_k$ at site $k$, and $\Gamma_k^{\sigma+}$ ($\Gamma_k^{\sigma-}$) is the new configuration when one electron with spin $\sigma$ is added to (removed from) the configuration $\Gamma_k$. For a half-filling system, $p(\Gamma_k)$ can be given by[1],

$$p(\Gamma_i)=\dfrac{\sum_{\Gamma_{Jj}}\xi^0_{\Gamma_{Ii},\Gamma_{Jj},\Gamma_{Ii},\Gamma_{Jj}}g(\Gamma_i)^2g(\Gamma_j)^2}{\sum_{\Gamma_{Ii},\Gamma_{Jj}}\xi^0_{\Gamma_{Ii},\Gamma_{Jj},\Gamma_{Ii},\Gamma_{Jj}}g(\Gamma_i)^2g(\Gamma_j)^2}\Bigg|_{\substack{n_{Ii\uparrow}+n_{Jj\uparrow}=2,\\n_{Ii\downarrow}+n_{Jj\downarrow}=2}} \quad (17)$$

The underlying physical meaning of Eq. (15) and (16) is discussed in detail in Ref. [10]. Eq. (15) is the key approximation to efficiently estimate the 1PDM and 2PCM that are required to evaluate the energy. It has been demonstrated to be more accurate than the commonly used GA for the single band Hubbard model, in both one- and two-dimension [10]. In this work, we use a 2-band Hubbard model with an extensive parameter space as a testbed to estimate the accuracy of Eq. (15) for multi-band/multi-orbital systems.

---

[1] For systems with other electron densities, the expression of $p(\Gamma_i)$ can be found in Ref. [10].

## 2.3. The rotationally invariant (RI) scheme.

In the original GCGM algorithm, the Gutzwiller projector is parameterized based on the onsite configurations $\Gamma$. As a result, the computational cost scales exponentially with the increasing number of electrons and orbitals, which can be a major bottleneck in the GCGM method. To reduce the computational complexity, a RI scheme is incorporated into the GCGM method to group certain orbitals and describe the system with the number of electrons occupying the orbitals instead of the configuration $\Gamma_i$. The Gutzwiller projector $\hat{G} = \prod_{Ii}\left(\sum_{\Gamma_{Ii}} g(\Gamma_i)|\Gamma_{Ii}\rangle\langle\Gamma_{Ii}|\right)$ in Eq. (7) has a diagonal form. If we use another orthonormal set of basis $\{|\tilde{\Gamma}_{Ii}\rangle\}$ instead of $\{|\Gamma_{Ii}\rangle\}$, the Gutzwiller projector becomes,

$$\hat{G} = \prod_{Ii}\left(\sum_{\Gamma_{Ii}} g(\Gamma_i)|\Gamma_{Ii}\rangle\langle\Gamma_{Ii}|\right) = \prod_{Ii}\left(\sum_{\Gamma_i} g(\Gamma_i)\sum_{\tilde{\Gamma}_{Ii}\tilde{\Gamma}'_{Ii}}|\tilde{\Gamma}_{Ii}\rangle\langle\tilde{\Gamma}_{Ii}|\Gamma_{Ii}\rangle\langle\Gamma_{Ii}|\tilde{\Gamma}'_{Ii}\rangle\langle\tilde{\Gamma}'_{Ii}|\right)$$
$$= \prod_{Ii}\sum_{\tilde{\Gamma}_{Ii}\tilde{\Gamma}'_{Ii}}\left(\sum_{\Gamma_{Ii}}\langle\tilde{\Gamma}_{Ii}|\Gamma_{Ii}\rangle g(\Gamma_i)\langle\Gamma_{Ii}|\tilde{\Gamma}'_{Ii}\rangle\right)|\tilde{\Gamma}_{Ii}\rangle\langle\tilde{\Gamma}'_{Ii}|. \quad (18)$$

If we want to enforce the diagonal form of $\hat{G}$ during the rotational transformation of the basis, we need the term $\sum_{\Gamma_{Ii}}\langle\tilde{\Gamma}_{Ii}|\Gamma_{Ii}\rangle g(\Gamma_i)\langle\Gamma_{Ii}|\tilde{\Gamma}'_{Ii}\rangle$ to be $g(\tilde{\Gamma}_i)\delta_{\tilde{\Gamma}_{Ii},\tilde{\Gamma}'_{Ii}}$. It is apparent that $\langle\tilde{\Gamma}_{Ii}|\Gamma_{Ii}\rangle$ and $\langle\Gamma_{Ii}|\tilde{\Gamma}'_{Ii}\rangle$ can be non-zero only when the electron number is conserved, .i.e. $n(\tilde{\Gamma}_{Ii},\sigma) = n(\Gamma_{Ii},\sigma)$ and $n(\Gamma_{Ii},\sigma) = n(\tilde{\Gamma}'_{Ii},\sigma)$, where $n(\Gamma_{Ii},\sigma)$ denotes the number of spin-$\sigma$ electrons of the configuration $\Gamma_{Ii}$. So, if $g$ is dependent only on $n(\Gamma_i,\sigma)$ instead of the configuration $\Gamma_i$ on site $i$, i.e. $g(\Gamma_i) = g(n(\Gamma_i,\sigma)) = g(n(\Gamma_i,\uparrow),n(\Gamma_i,\downarrow))$, the term in Eq. (18) becomes

$$\sum_{\Gamma_{Ii}} \langle \tilde{\Gamma}_{Ii} | \Gamma_{Ii} \rangle g(\Gamma_i) \langle \Gamma_{Ii} | \tilde{\Gamma}'_{Ii} \rangle = g(n(\Gamma_i), \sigma) \sum_{\Gamma_{Ii}} \langle \tilde{\Gamma}_{Ii} | \Gamma_{Ii} \rangle \langle \Gamma_{Ii} | \tilde{\Gamma}'_{Ii} \rangle = g(n(\Gamma_i), \sigma) \delta_{\tilde{\Gamma}_{Ii}, \tilde{\Gamma}'_{Ii}}$$. So, Eq. (18) becomes,

$$\hat{G} = \prod_{Ii} \sum_{\tilde{\Gamma}_{Ii}} g(n(\tilde{\Gamma}_i, \sigma)) | \tilde{\Gamma}_{Ii} \rangle \langle \tilde{\Gamma}_{Ii} |, \tag{19}$$

which satisfies the diagonal form. Eq. (19) indicates that the system should be parameterized based on $n(\Gamma_i, \sigma)$ instead of $\Gamma_i$ if we want to keep the off-diagonal term of $\hat{G}$ to be zero during the rotational transformation of the basis. So, we can group the orbitals if they satisfy certain symmetry and use the number of electrons occupying the grouped orbitals to describe the system. The advantage of doing this is that the number of possible configurations grows exponentially with orbitals while the possible number of electrons occupying the grouped orbitals grows only linearly.

In the RI scheme, the 1PDM in Eq. (9) can be rewritten as,

$$\langle c^\dagger_{Ii\alpha\sigma} c_{Ii\beta\sigma} \rangle_{GWF} = \frac{1}{\langle \Psi_{GWF} | \Psi_{GWF} \rangle_{Ii,Jj}} \sum_{\substack{\{n_{Ii\uparrow}\},\{n_{Ii\downarrow}\} \\ \{n_{Jj\uparrow}\},\{n_{Jj\downarrow}\}}} g(\{...,n_i(A,\sigma), n_i(B,\sigma),...\}, \{n_i(-\sigma)\}) \cdot$$
$$g(\{...,n_i(A,\sigma)-1, n_i(B,\sigma)+1,...\}, \{n_i(-\sigma)\}) g(\{n_j(\uparrow)\}, \{n_j(\downarrow)\})^2 \cdot$$
$$F_\uparrow(n_{Ii\uparrow}, n_{Jj\uparrow}) F_\downarrow(n_{Ii\downarrow}, n_{Jj\downarrow}) \sum_{\Gamma_{Ii},\Gamma'_{Ii},\Gamma_{Jj}} \langle \Gamma_{Ii} | c^\dagger_{Ii\alpha\sigma} c_{Ii\beta\sigma} | \Gamma'_{Ii} \rangle \xi^0_{\Gamma_{Ii},\Gamma_{Jj},\Gamma'_{Ii},\Gamma_{Jj}},$$

(20)

when orbital $\alpha$ and $\beta$ cannot be grouped together, i.e. $\alpha$ belongs to group $A$ of orbitals and $\beta$ belongs to a different group $B$. $-\sigma$ denotes the opposite spin of $\sigma$. Here, $g(\Gamma_j)$ in Eq. (9) is now rewritten as $g(\{n_j(\uparrow)\}, \{n_j(\downarrow)\})$, $g(\Gamma_i)$ rewritten as $g(\{...,n_i(A,\sigma), n_i(B,\sigma),...\}, \{n_i(-\sigma)\})$, $g(\Gamma'_i)$ rewritten as $g(\{...,n_i(A,\sigma)-1, n_i(B,\sigma)+1,...\}, \{n_i(-\sigma)\})$. Here, $\Gamma_{Ii}$ and $\Gamma'_{Ii}$ are correlated by the selection rule: $\langle \Gamma_{Ii} | c^{\dagger}_{Ii\alpha\sigma} c_{Ii\beta\sigma} | \Gamma'_{Ii} \rangle$ in Eq. (20) is non-zero only if $\Gamma_{Ii}$ and $\Gamma'_{Ii}$ differ by one spin orbital (one electron hops from orbital $\alpha$ to $\beta$). For $\Gamma_{Ii}$, if the number of electrons occupying orbital group $(A,\sigma)$ and $(B,\sigma)$ are $n_i(A,\sigma)$ and $n_i(B,\sigma)$, then for $\Gamma'_{Ii}$ the number of electrons occupying orbital group $(A,\sigma)$ and $(B,\sigma)$ are $n_i(A,\sigma)-1$ and $n_i(B,\sigma)+1$, respectively. After orbitals are grouped under the RI principle, the $g$-factors can also be grouped, and those common $g$-factors can be factored out from the summation in Eq. (9), which then becomes Eq. (20). When orbital $\alpha$ and $\beta$ can be grouped in the same group $A$, Eq. (9) can be rewritten as,

$$\langle c^{\dagger}_{Ii\alpha\sigma} c_{Ii\beta\sigma} \rangle_{GWF} = \frac{1}{\langle \Psi_{GWF} | \Psi_{GWF} \rangle_{Ii,Jj}} \sum_{\substack{\{n_{Ii\uparrow}\},\{n_{Ii\downarrow}\} \\ \{n_{Jj\uparrow}\},\{n_{Jj\downarrow}\}}} g\big(\{n_i(\uparrow)\},\{n_i(\downarrow)\}\big)^2 \cdot$$
$$g\big(\{n_j(\uparrow)\},\{n_j(\downarrow)\}\big)^2 F_{\uparrow}(n_{Ii\uparrow},n_{Jj\uparrow}) F_{\downarrow}(n_{Ii\downarrow},n_{Jj\downarrow}) \cdot \quad (21)$$
$$\sum_{\Gamma_{Ii},\Gamma'_{Ii},\Gamma_{Jj}} \langle \Gamma_{Ii} | c^{\dagger}_{Ii\alpha\sigma} c_{Ii\beta\sigma} | \Gamma'_{Ii} \rangle \xi^0_{\Gamma_{Ii},\Gamma_{Jj},\Gamma'_{Ii},\Gamma_{Jj}},$$

where both $g(\Gamma_i)$ and $g(\Gamma'_i)$ in Eq. (9) can be now rewritten as $g\big(\{...,n_i(A,\sigma),...\},\{n_i(-\sigma)\}\big)$, i.e. $g\big(\{n_i(\uparrow)\},\{n_i(\downarrow)\}\big)$, since the number of electrons occupying the orbitals of group $A$ are the same for $\Gamma_i$ and $\Gamma'_i$. Eq. (10) can be rewritten as,

$$\langle c^{\dagger}_{Ii\alpha\sigma} c_{Jj\beta\sigma} \rangle_{GWF} = \frac{1}{\langle \Psi_{GWF} | \Psi_{GWF} \rangle_{Ii,Jj}} \sum_{\substack{\{n_{Ii\uparrow}\},\{n_{Ii\downarrow}\} \\ \{n_{Jj\uparrow}\},\{n_{Jj\downarrow}\}}} g\big(\{...,n_i(A,\sigma),...\},\{n_i(-\sigma)\}\big) \cdot$$
$$g\big(\{...,n_j(B,\sigma),...\},\{n_j(-\sigma)\}\big) g\big(\{...,n_i(A,\sigma)-1,...\},\{n_i(-\sigma)\}\big) \cdot \quad (22)$$
$$g\big(\{...,n_j(B,\sigma)+1,...\},\{n_j(-\sigma)\}\big) F_{\uparrow}(n_{Ii\uparrow},n_{Jj\uparrow}) F_{\downarrow}(n_{Ii\downarrow},n_{Jj\downarrow}) \cdot$$
$$\sum_{\Gamma_{Ii},\Gamma'_{Ii},\Gamma_{Jj},\Gamma'_{Jj}} \langle \Gamma_{Ii} | c^{\dagger}_{Ii\alpha\sigma} | \Gamma'_{Ii} \rangle \langle \Gamma_{Jj} | c_{Jj\beta\sigma} | \Gamma'_{Jj} \rangle \xi^0_{\Gamma_{Ii},\Gamma_{Jj},\Gamma'_{Ii},\Gamma'_{Jj}}, \; for\;(I,i)\neq(J,j)$$

The expression of the onsite 2PCM $\langle c_{Ii\alpha\sigma}^{\dagger} c_{Ii\beta\sigma}^{\dagger} c_{Ii\delta\sigma} c_{Ii\gamma\sigma} \rangle_{GWF}$ has a similar form with Eq. (20)(21) and is not presented here for conciseness. Eq. (12) becomes,

$$\langle \Psi_{GWF} | \Psi_{GWF} \rangle_{Ii,Jj} = \sum_{\substack{\{n_{Ii\uparrow}\},\{n_{Ii\downarrow}\} \\ \{n_{Jj\uparrow}\},\{n_{Jj\downarrow}\}}} g(\{n_i(\uparrow)\},\{n_i(\downarrow)\})^2 g(\{n_j(\uparrow)\},\{n_j(\downarrow)\})^2 \cdot F_{\uparrow}(n_{Ii\uparrow}, n_{Jj\uparrow}) F_{\downarrow}(n_{Ii\downarrow}, n_{Jj\downarrow}) \sum_{\Gamma_{Ii},\Gamma_{Jj}} \xi^0_{\Gamma_{Ii},\Gamma_{Jj},\Gamma_{Ii},\Gamma_{Jj}} \quad (23)$$

By using the RI approach, the computational cost will be significantly reduced. If we compare, for example, Eq. (9) and Eq. (20), we can find that the summation in Eq. (9) goes over all the possible configurations $\{\Gamma_{Ii}, \Gamma_{Jj}\}$ ($\Gamma'_{Ii}$ are determined by $\Gamma_{Ii}$ with the selection rule) while the summation in Eq. (20) breaks down into 2 parts: the 1st summation goes over all the possible electron numbers occupying the grouped orbitals $\{n_{Ii}(\uparrow), n_{Ii}(\downarrow), n_{Jj}(\uparrow), n_{Jj}(\downarrow)\}$ and the 2nd summation goes over all the possible configurations $\{\Gamma_{Ii}, \Gamma_{Jj}\}$, of which the electron numbers occupying the grouped orbitals satisfy $\{n_{Ii}(\uparrow), n_{Ii}(\downarrow), n_{Jj}(\uparrow), n_{Jj}(\downarrow)\}$. Since all included in the 2nd summation are the selection rule and the predetermined HF coefficients $\xi^0_{\Gamma_{Ii},\Gamma_{Jj},\Gamma_{Ii},\Gamma_{Jj}}$, the summations can be evaluated only once and be stored in the memory as invariants that are not dependent on $g$-factors. During the minimization of energy, we only need to evaluate the 1st summation over $\{n_{Ii}(\uparrow), n_{Ii}(\downarrow), n_{Jj}(\uparrow), n_{Jj}(\downarrow)\}$, the possible choices of which scales only linearly with the number of orbitals. Therefore, the computational cost is greatly reduced compared to the cost of evaluating the summation over $\{\Gamma_{Ii}, \Gamma_{Jj}\}$ in Eq. (9), the possible choices of which scales exponentially with the number of orbitals.

## 3. RESULTS

In this section we test our GCGM approach with the 1D 2-band Hubbard model described by Eq. (1)-(5) and compare our numerical results with those given by the DMRG method as implemented in the ITensor library [21]. In DMRG calculations, a large system containing 100 sites with open boundaries is used to ensure that the bulk properties are approached at the center. For the 2-band model at half-filling, each site contains 2 electrons with opposite spins in the initial matrix product state. Observables such as the double occupancy are averaged over the central 20 sites.

Firstly, we focus on the original GCGM method without the RI add-on to validate the GCGM method itself. We systematically compute the ground-state energy per site and the double occupancy. We then continue to test the accuracy of the RI approach, which is integrated into the GCGM method to largely improve its efficiency. The electron density is 2 electron per site, i.e., half-filling throughout this work. All the calculations are at zero temperature. To conclude our benchmark test, we study a Mott transition between a metal and a Mott insulator by evaluating the charge gap. The phase diagram given by GCGM with RI agrees reasonably well with that given by DMRG.

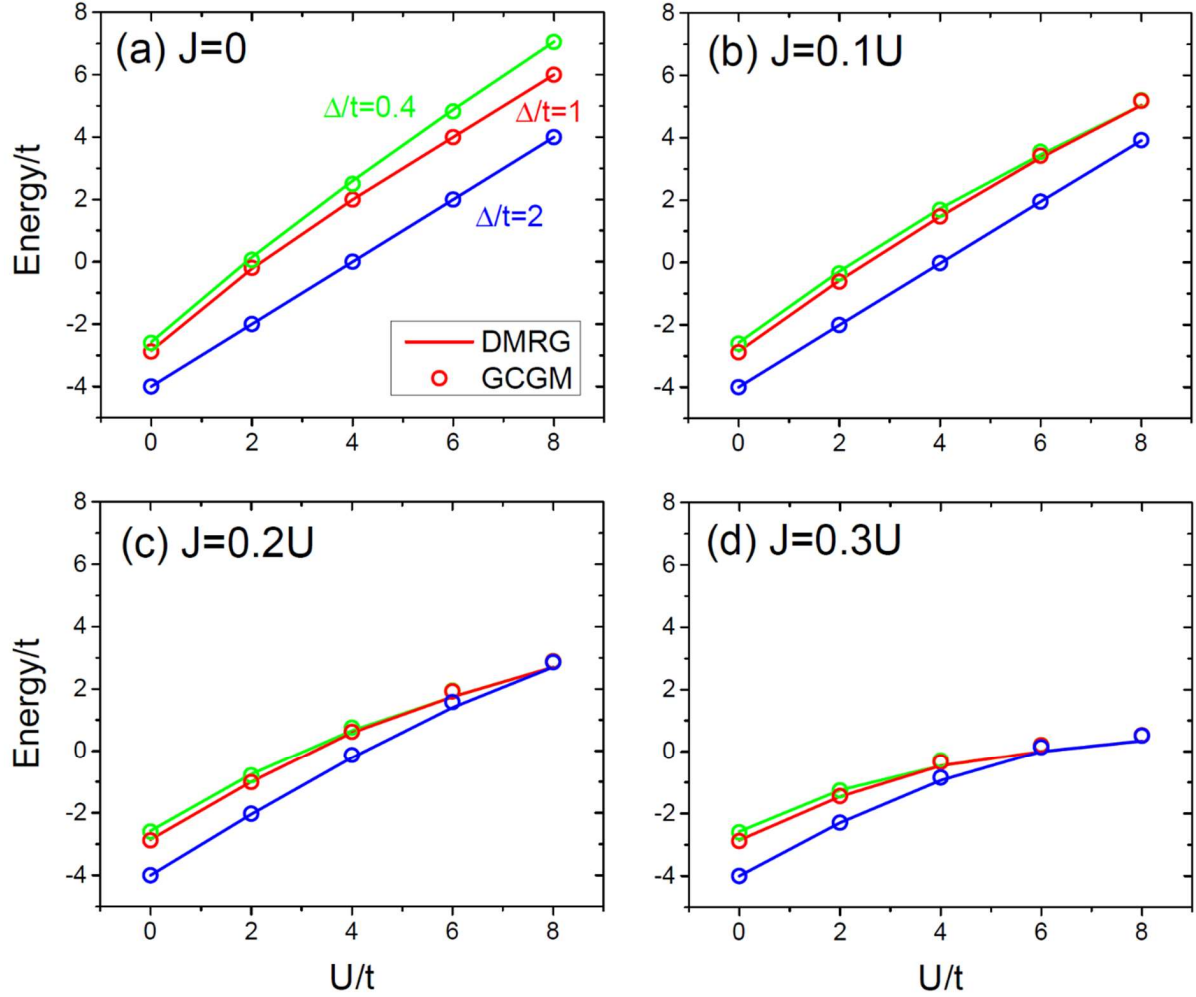

Figure 1. The ground state energy of the 1D 2-band Hubbard model given by the GCGM and the DMRG method as a function of $U/t$ with (a) $J=0$, (b) $J=0.1U$, (c) $J=0.2U$, and (d) $J=0.3U$; and with the crystal-field splitting $\Delta = 0.4t, t, 2t$. The constraint $U = U' + 2J$ is imposed.

Figure 1 plots the ground state energy as a function of $U/t$ given by the original GCGM method and the DMRG method. In the GCGM calculation 32 atomic sites, or equivalently, 32 k-points are used; while in the DMRG calculation 100 atomic sites are used. The periodic boundary condition is not enforced in the DMRG calculation. Thus, more atomic sites are needed to simulate a periodic system. We perform the test over a wide range of $J$, the Hund coupling, and

$\Delta\left(=\Delta_2=-\Delta_1=t'\right)$, the crystal-field splitting. We present the results with selected values of $J = 0, 0.1U, 0.2U, 0.3U$ and $\Delta = 0.4t, t, 2t$. For smaller $J$ of 0 or $0.1U$, the GCGM gives results in satisfying agreement with the DMRG method. For larger $J$ of $0.2U$ or $0.3U$, the GCGM gives energies that are slightly higher than the DMRG energies at intermediate to strong interaction strengths. Other than that, the GCGM method in general estimates the energy accurately. The two methods give an energy difference $\varepsilon_{Energy} \leq 0.21t$. Figure 2 plots the double occupancy (defined as $\langle n_{I\uparrow} n_{I\downarrow} \rangle_{GWF} = \langle c^\dagger_{I\uparrow} c_{I\uparrow} c^\dagger_{I\downarrow} c_{I\downarrow} \rangle_{GWF}$) of the ground state as a function of $U/t$ given by the original GCGM method and the DMRG method. The GCGM method, in general, gives a qualitatively correct estimate of the double occupancy. It sometimes moderately overestimates the double occupancy compared to the DMRG results. The most obvious discrepancy happens at the intermediate to strong correlation strength ($4 \leq U/t \leq 6$). The two methods give a difference of double occupancy $\varepsilon_D \leq 0.038$. We note that the evaluation of the double occupancy is not as accurate as that in the single-band case [10].

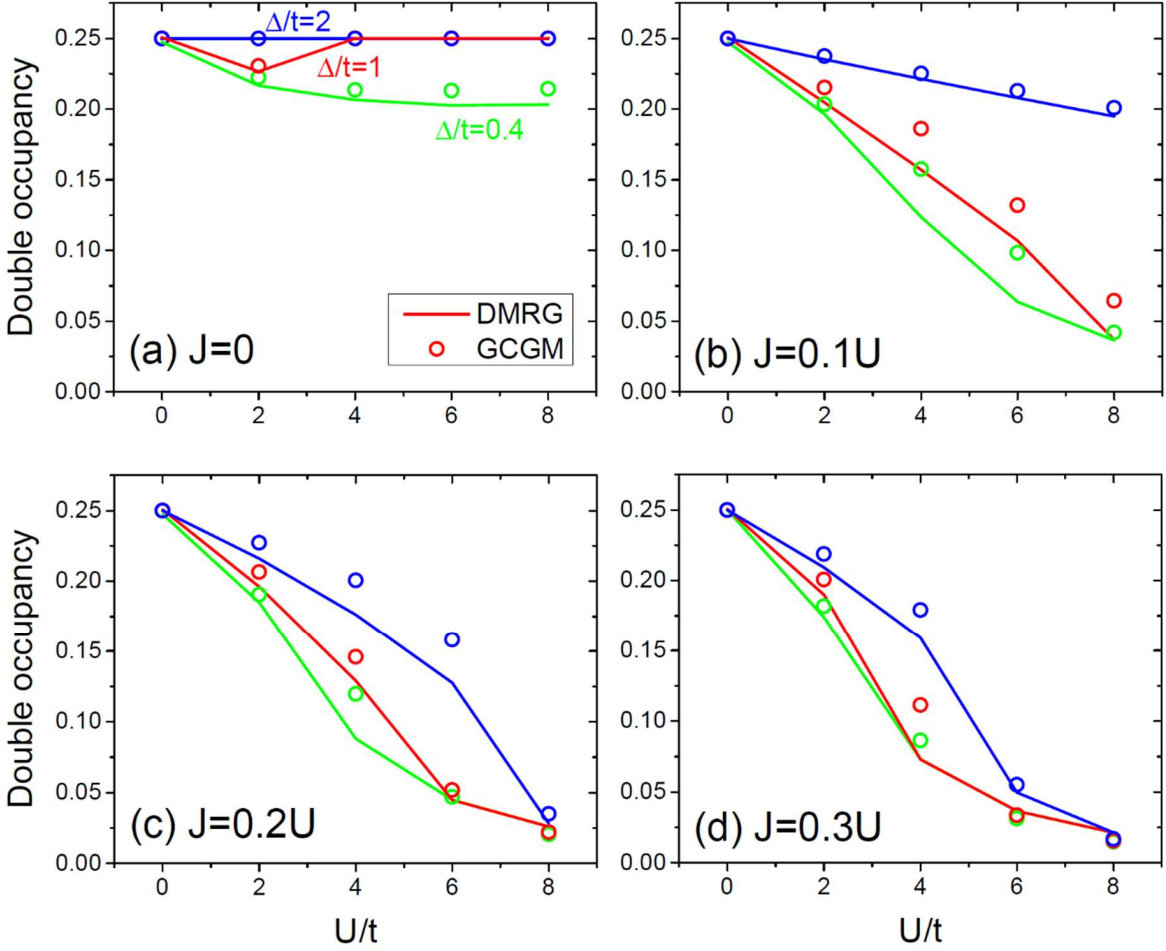

Figure 2. The double occupancy of the 1D 2-band Hubbard model given by the GCGM and the DMRG method as a function of $U/t$ with (a) $J=0$, (b) $J=0.1U$, (c) $J=0.2U$, and (d) $J=0.3U$; and with the crystal-field splitting $\Delta = 0.4t, t, 2t$.

Figures 3 and 4 plot the ground state energy and the double occupancy, respectively, as a function of $U/t$ given by the original GCGM method and the GCGM within the RI scheme. The RI approximation is accurate in the limit of $J=0$, as verified in the calculations of the energy (Fig. 3(a)) and of the double occupancy (Fig. 4(a)). When $J \neq 0$, using the RI approach still nearly perfectly reproduces the energy. The discrepancy in the energies given by the 2 methods is small with $\varepsilon_{Energy} \leq 0.08$, demonstrating that the RI approach is a good approximation in describing the energy. However, the RI approach introduces some inaccuracy when it is used to estimate the double occupancy, as shown in Fig. 4(b)-(d). It often overestimates the double occupancy, particularly at the weak to intermediate correlation strength. The most obvious discrepancy ($\varepsilon_D \approx 0.049$) happens when $J = 0.1U$, $\Delta = 2t$, as shown in Fig. 4(b). The minimized energy favors a pair of electrons of opposite spin occupying both orbitals (e.g., the spin-up electron occupies orbital #1 and the spin-down electron occupies orbital #2) of instead only 1 orbital (e.g., both electrons occupy orbital #1). So, the $g$-factor of the $1^{st}$ case is larger than that of the $2^{nd}$ case. However, in the rotationally invariant approximation, the two scenarios are treated equally and the $g$-factors are enforced to be the same for the two scenarios, resulting in the discrepancy as seen in Fig. 4(b). Nevertheless, although the estimation of the double occupancy is not very accurate with the rotationally invariant approach, the energy is well reproduced, as shown in Fig. 3(b).

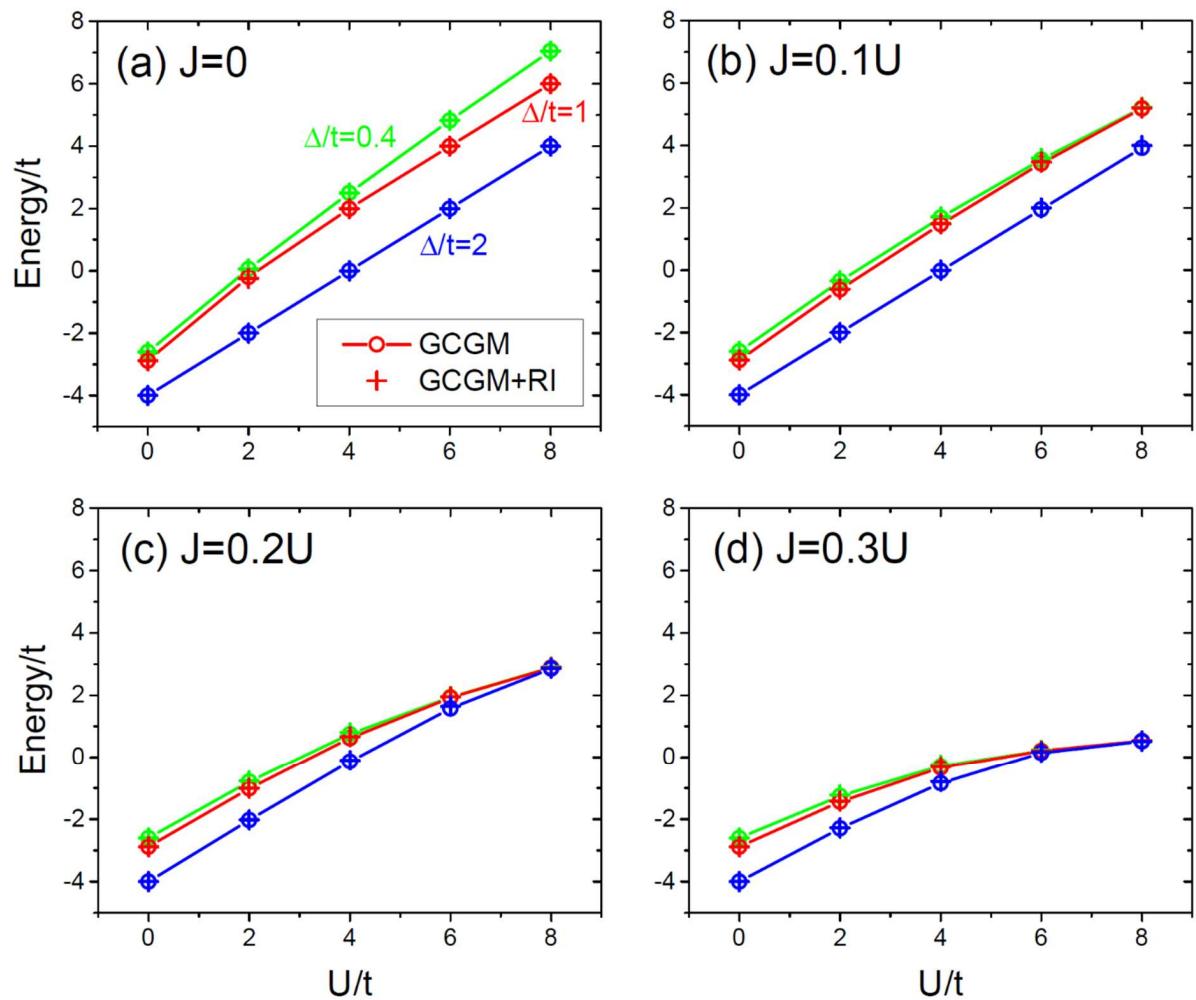

Figure 3. The ground state energy of the 1D 2-band Hubbard model given by the original GCGM method and the GCGM with the rotationally invariant (RI) approach as a function of $U/t$ with (a) $J=0$, (b) $J=0.1U$, (c) $J=0.2U$, and (d) $J=0.3U$; and with the crystal-field splitting $\Delta = 0.4t, t, 2t$.

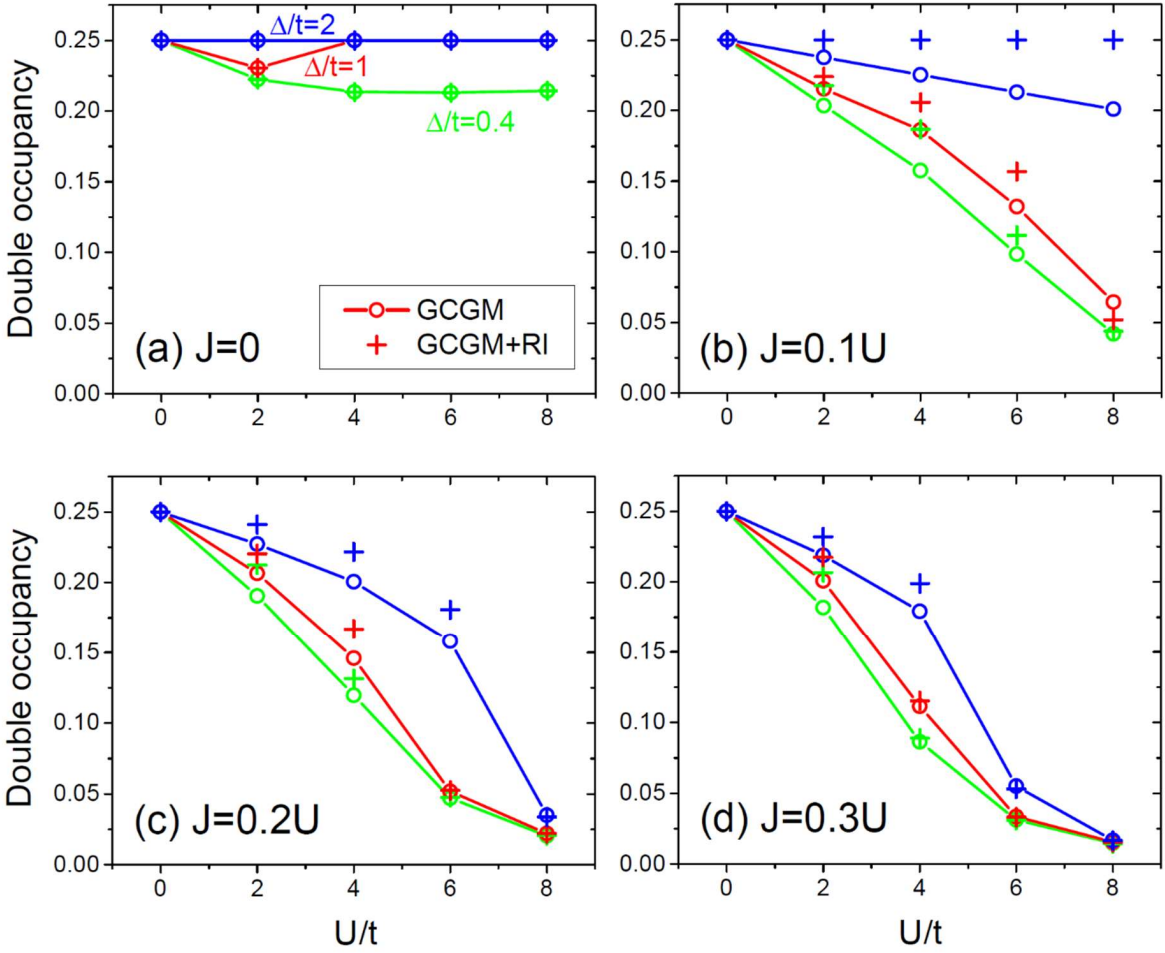

Figure 4. The double occupancy of the 1D 2-band Hubbard model given by the original GCGM method and the GCGM with the RI approach as a function of $U/t$ with (a) $J=0$, (b) $J=0.1U$, (c) $J=0.2U$, and (d) $J=0.3U$; and with the crystal-field splitting $\Delta = 0.4t, t, 2t$.

An original motivation for the Hubbard model is to use it as a prototype to achieve fundamental understanding of the Mott transition between a metal and a Mott insulator induced by the repulsive interactions among electrons [22-26]. For the single-band 1D Hubbard model, Lieb and Wu established that Mott transition does not exist and the system is insulating at any finite $U$ [27]. In contrast, the 2-band Hubbard model displays a rich behavior of Mott transitions [28,29], thus providing another testbed to benchmark the GCGM method.

We use the charge gap $\Delta_c$ to identify a Mott transition. $\Delta_c$ can be defined as

$$\Delta_c = \frac{1}{2}\left[E(N/2+1, N/2+1) + E(N/2-1, N/2-1) - 2E(N/2, N/2)\right], \quad (24)$$

where $E(x, y)$ is the ground state energy of the system with $x$ spin-up electrons and $y$ spin-down electrons. $\Delta_c$ should be zero in a metallic state and takes finite values when the system becomes insulating. However, it is numerically challenging to capture a vanishing $\Delta_c$ according to Eq. (24) since it involves obtaining a small number by subtraction of large numbers. To see this, we show in Fig. 5 $\Delta_c$ as a function of $U$ for the crystal-field splitting $\Delta = 0.6t$ and the Hund coupling $J = 0.2U$, calculated by both DMRG and GCGM. The inset in Fig. 5 gives the band structure of the non-interacting Hamiltonian ($U = 0$), in which the two bands are separated by $\delta = 2\Delta$. As long as $\delta < 4t$ (the bandwidth), or $\Delta < 2t$, the non-interacting system is metallic. However, as shown in Fig. 5, both DMRG and GCGM give small (less than 1% of the total energy) $\Delta_c$ when $U$ is close to zero, due to numerical inaccuracy of the methods. Nevertheless, one can clearly identify a "kink" at $U \approx 3.6t$ for both DMRG and GCGM results, which shows a sudden change from weak to strong response of $\Delta_c$ with respect to $U$, signaling a metal-to-insulator transition.

Similar calculations are repeated for a range of $\Delta$ values to map out the critical $U$ at the transition state ($U_c$) as a function of $\Delta$, which is plotted in Fig. 6. Compared with the DMRG results, GCGM overestimates $U_c$ for relatively small $\Delta$ and underestimates $U_c$ for large $\Delta$. Overall, the GCGM results agree reasonably well with those derived by DMRG, and the agreement appears to be better with stronger Hund's coupling ($J = 0.2U$).

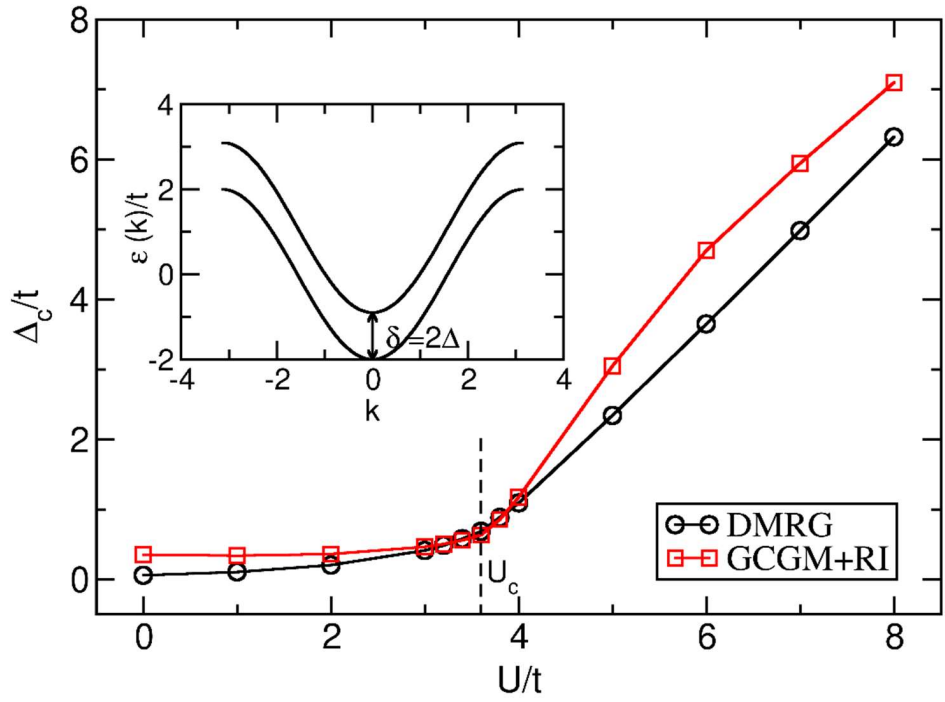

Figure 5. $\Delta_c$ as a function of $U$ calculated by DMRG and GCGM with rotational invariance. Inset gives the band structure of the non-interacting Hamiltonian.

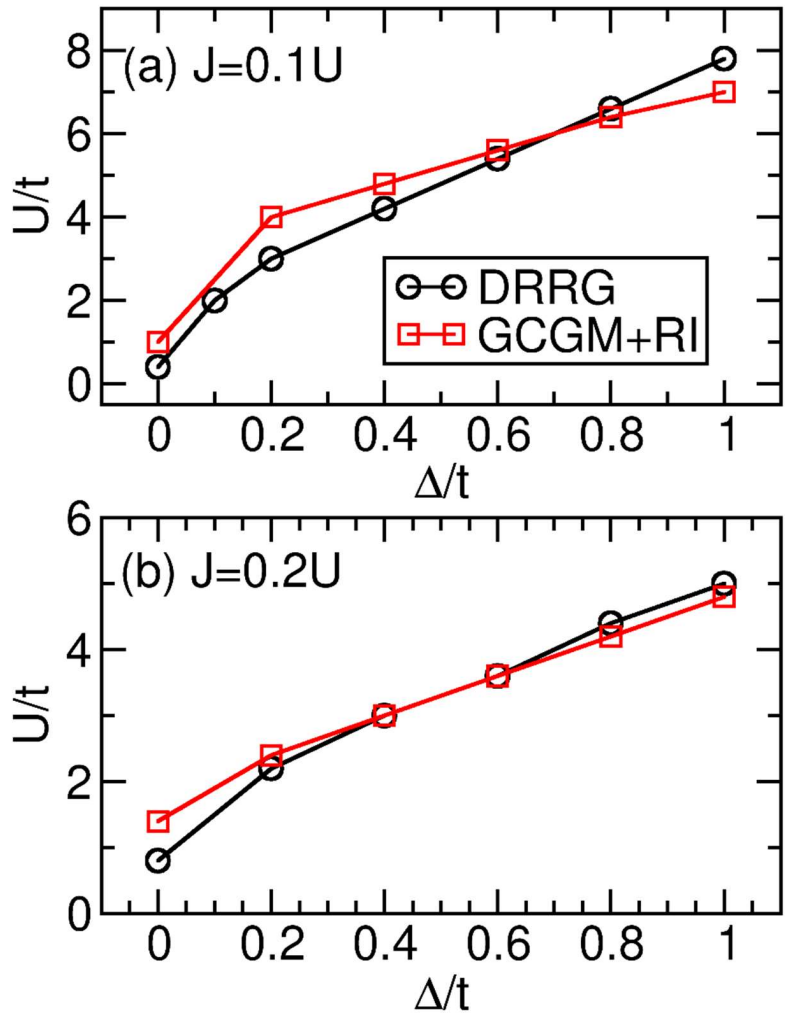

Figure 6. The critical $U$ for metal-to-Mott insulator transitions as a function of the crystal-field splitting $\Delta/t$ for (a) $J = 0.1U$ and (b) $J = 0.2U$.

Now we discuss the speedup of the RI approach. Table 1 lists the computation time of the 3 methods. The original GCGM and GCGM with RI were run on a single core since they are efficient enough, while a parallel version of the DMRG code was executed on one computing node with 32 cores. In the GCGM calculation 32 atomic sites are used with periodic boundary conditions; while in the DMRG calculation 100 atomic sites are used without periodic boundary conditions. The original GCGM takes ~12 seconds with one core to calculate and store the predetermined factors ($\xi^0_{\Gamma_{Ii},\Gamma_{Jj},\Gamma'_{Ii},\Gamma'_{Jj}}$ in Eq. (14)) and ~20 seconds per iteration during the minimization of the energy. It usually takes up to 30 iterations for the energy to converge. The GCGM in conjunction with the RI takes ~11 seconds to compute and store the prefactors (the summations that contain $\xi^0_{\Gamma_{Ii},\Gamma_{Jj},\Gamma'_{Ii},\Gamma'_{Jj}}$ in Eq.(20)-(23)) and the energy usually converges within $1 \sim 2$ seconds, that is, $0.04 \sim 0.06$ seconds roughly per iteration. The predetermined factors need to be estimated only once and do not change in the process of energy minimization. The speedup is $300 \sim 500\times$ for each iteration, and the overall speedup is $\sim 50\times$ if it takes $\sim 30$ iterations to converge. Besides, the GCGM with RI requires a reduced memory to store all the predetermined coefficients from $|\Psi_0\rangle$. For the original GCGM, there are about 157,000 $\xi^0_{\Gamma_{Ii},\Gamma_{Jj},\Gamma'_{Ii},\Gamma'_{Jj}}$ that need to be pre-calculated and stored. For the GCGM in conjunction with RI, we only need to calculate and store ~13,000 prefactors since $\xi^0_{\Gamma_{Ii},\Gamma_{Jj},\Gamma'_{Ii},\Gamma'_{Jj}}$ can be grouped according to the electron occupancies of the configurations. That is a ~10 folds reduction from the original GCGM. DMRG, on the other hand, is a much more expensive method. It takes about 1 h·core to complete an iteration in DMRG, compared with 0.04-0.06 s·core in GCGM+RI. Considering the apparent slower convergence (more iterations) of GCGM+RI than DMRG and the overhead cost in prefactor calculations, the GCGM+RI method is still 3 orders of magnitude more efficient than DMRG.

| Method | Original GCGM | GCGM + RI | DMRG |
|---|---|---|---|
| Time (s·core) to compute pre-factors | 12 | 11 | NA |
| Time (s·core) for each iteration/sweep | 20 | 0.04-0.06 | 3,600 |
| Number of iterations/sweeps | 30 | 30 | 10 |
| Total computation time (s·core) | 612 | 13 | 36,000 |

Table 1. The computation time of the 3 methods with a single-core: the original GCGM, GCGM with RI and DMRG.

We can roughly estimate the computational complexity of the GCGM with/without the use of RI by considering a general single atom per cell $N$-orbital (i.e. $2N$ spin orbitals) system, where the $N$ orbitals can be grouped under the RI scheme. To estimate nearest neighbor hopping $\langle c_{Ii\alpha\sigma}^{\dagger} c_{Jj\beta\sigma} \rangle_{GWF}$ or the 2-particle term $\langle c_{Ii\alpha\sigma}^{\dagger} c_{Ii\beta\sigma'}^{\dagger} c_{Ii\delta\sigma'} c_{Ii\gamma\sigma} \rangle_{GWF}$, we need to sum over all the possible configurations $(\Gamma_{Ii}, \Gamma_{Jj})$ without the use of RI. There are $2^{2N}$ choices for each of $(\Gamma_{Ii}, \Gamma_{Jj})$, $2^{2N} \times 2^{2N} = 2^{4N}$ possible combinations of $(\Gamma_{Ii}, \Gamma_{Jj})$. So, the complexity to compute the total energy $E_{GWF}$ is proportional to $2^{4N}$. The complexity to compute the derivative for each $\frac{\partial E_{GWF}}{\partial g(\Gamma_i)}$ is also proportional to $2^{4N}$. Given there are altogether $2^{2N}$ $g$-factors, the complexity to compute all the derivatives is proportional to $2^{2N} \times 2^{4N} = 2^{6N}$. With the use of RI, the summation would be over the number of spin-up and spin-down electrons occupying the grouped $N$ orbitals $\{n_{Ii}(\uparrow), n_{Ii}(\downarrow), n_{Jj}(\uparrow), n_{Jj}(\downarrow)\}$. For example, $n_{Ii}(\uparrow)$, the number of spin-up electrons occupying the grouped $N$ orbitals, could be $0, 1, 2, ..., N$, i.e. $N+1$ choices. Therefore, there are $(N+1)^4$ choices of $\{n_{Ii}(\uparrow), n_{Ii}(\downarrow), n_{Jj}(\uparrow), n_{Jj}(\downarrow)\}$. The complexity to compute $E_{GWF}$ is thus proportional to $(N+1)^4$. As there are $(N+1)^2$ $g$-factors, the complexity to compute all the derivatives $\frac{\partial E_{GWF}}{\partial g(\Gamma_i)}$ is proportional to $(N+1)^4 \times (N+1)^2 = (N+1)^6$. The scaling is now reduced from exponential to polynomial with the use of RI.

In the above rough estimate of complexity, we overlook some factors. For example, the terms of selection rule (such as $\langle \Gamma_{Ii} | c^\dagger_{Ii\alpha\sigma} c_{Ii\beta\sigma} | \Gamma'_{Ii} \rangle$ in Eq. (9)) are now packed into the predetermined coefficients with the use of RI and need to be evaluated only once, resulting in further reduced complexity. The electron density also plays a role in the evaluation of the predetermined coefficients. For example, half-filling often gives the largest number of non-zero coefficients. But since these coefficients are evaluated only once, the electron density likely has little impact on the reduction of complexity with the use of RI.

4. SUMMARY

In our previous studies, we have developed the GCGM method for correlated electron systems and benchmarked it with molecules [8,9,12] and bulk systems, in particular, the Hubbard model [10,11]. As our previous study only focused on the single-band Hubbard model, in this work we continue our investigation and validate the GCGM method with a 2-band/orbital Hubbard model with demonstrated accuracy. One fundamental challenge of solving the many-body Schrödinger equation for multi-orbital systems is that the computational effort grows exponentially with the number of onsite orbitals. To overcome this bottleneck, we introduce a RI approach to efficiently group orbitals that satisfy certain symmetries and parametrize the system based on the number of electrons occupying the grouped orbitals. We validate the RI approach with ground-state energy calculations and characterizations of a Mott transition between a metal and a Mott insulator. A $\sim 50\times$ speedup makes GCGM in conjunction with RI much faster than the original GCGM and DMRG.

Ref. [12] and this work provide a systematic validation of the RI approach on smaller molecular and bulk systems. With the promising speedup of the RI approach, we will be able to study larger systems with GCGM. Multi-orbital molecules for which accurate data are available for comparison (e.g., $Cr_2$) are a good test bed where the performance of GCGM on real materials can be evaluated to help the development of the method towards treating real correlated bulk materials. Another benchmark study would be the 3-band or 4-band Hubbard models. After these benchmark studies of more complex molecules and bulk systems are accomplished, calculations of large, realistic bulk systems will be readily carried out.

ACKNOWLEDGES

Work at Ames National Laboratory was supported by the US Department of Energy (DOE), Office of Science, Basic Energy Sciences, Materials Science and Engineering Division including a grant of computer time at the National Energy Research Scientific Computing Centre (NERSC) in Berkeley. Ames Laboratory is operated for the US DOE by Iowa State University under Contract No. DE-AC02-07CH11358.